\documentclass[showpacs,amssymb,aps,amsmath,twocolumn]{revtex4}
\usepackage{epsfig}

\newcommand{\tr}{t_{\rm rel}}
\newcommand{\upd}{{\rm d}}

\begin{document}

\title{Non-topographic description of inherent structure dynamics in
glass formers}

\author{Ludovic Berthier$^{1,2}$ and Juan P. Garrahan$^1$}

\affiliation{$^1$Theoretical Physics, University of Oxford, 1 Keble
Road, Oxford, OX1 3NP, UK \\ $^2$Laboratoire des Verres, Universit\'e
Montpellier II, 34095 Montpellier, France}

\date{\today}

\begin{abstract}
We show that the dynamics between inherent structures in glass forming
systems can be understood in purely dynamical terms, without any
reference to ``topographic'' features of the potential energy
landscape.  This ``non-topographic'' interpretation is based instead
on the existence of dynamical heterogeneities and on their statistical
properties. Our view is supported by the study of simple dynamically
facilitated models of glass formers. These models also allow for the
formulation of quantitative theoretical predictions which are
successfully compared to published data obtained in numerical and
experimental studies of local dynamics of supercooled liquids.
\end{abstract}


\maketitle

{\small \it It's been like a kind of capsule, a bubble in time and
space...}

{\small \hfill D. Lodge, \it Paradise News}

\section{Introduction}

In 1969, Goldstein suggested that the dynamical slowdown of a liquid
approaching its glass transition could be understood in terms of its
potential energy landscape~\cite{goldstein}. The practical
implementation of this description was later introduced by Stillinger
and Weber via the inherent structure (IS) formalism~\cite{SW1}. The
landscape or ``topographic'' view has become one of the paradigms in
the field~\cite{topo,Debe}. The purpose of this paper is to pursue,
however, the rather orthogonal idea that the landscape description is
essentially incomplete, in the sense that from the statistical
properties of IS, or similar static characteristics, it is not
possible to capture the central spatial and dynamical aspects of the
physics of glass formers. Instead, we aim here at a
``non-topographic'' reinterpretation of the dynamics between IS,
showing that the central role is played by dynamical heterogeneities,
without resorting to any special features of the potential energy
landscape.

The connection between the existence of a dynamical correlation length
and the landscape description is in fact implicitly present in the
original papers~\cite{goldstein,SW2}, although it only clearly appears
in more recent
works~\cite{Buchner-Heuer2,Buchner-Heuer,Denny-et-al,Doliwa-Heuer,dh2}.
The connection relies on two remarks. (i) According to
Goldstein~\cite{goldstein}, the landscape is useful for ``viscous''
liquids, for which a separation of time scales between fast vibrations
and slower structural relaxations exists. 
This gives rise to the notions of ``basins'' or ``traps'' 
which are then purely dynamical concepts.
(ii) Escape
from these traps was described to be ``localized'' in
space~\cite{goldstein}, a fact which was later numerically
confirmed~\cite{SW2}.  The transition rate between basins
therefore
becomes extensive ``for large enough systems''~\cite{SW2}.  How large
the system must be, what is the geometry, and what are the typical
length and time scales of these rearrangements are crucial
questions left unanswered unless spatial and dynamical aspects are
included explicitly in the theoretical description.

In what follows we show that natural answers
can be obtained in the context of dynamical
heterogeneities~\cite{review}, thus leading to a consistent
qualitative interpretation of IS dynamics. This we demonstrate using
the simplest models which capture the phenomenon of dynamic
heterogeneity, and for which the landscape properties are completely
irrelevant (sections II and III). We also show that a satisfactory
quantitative agreement with numerical and experimental studies of
mesoscopic systems can be obtained (section IV), emphasizing the
relevance of the non-topographic description of supercooled liquid
dynamics discussed in this paper.

\section{Dynamical heterogeneities versus ``hidden inherent 
structures''}

In a cold dense liquid mobility is sparse. A microscopic immobile
region is mostly surrounded by other immobile regions. It can become
mobile only if it is next to a rare microscopic mobile region: its
dynamics is facilitated by the presence of an unjammed mobile
neighbour. This is the concept of dynamic facilitation, originally
introduced in~\cite{Fredrickson-Andersen}, which leads to the
phenomenon of dynamic heterogeneity~\cite{harro,Garrahan-Chandler}.

The elementary models which capture the idea of dynamic facilitation
are the Fredrickson-Andersen (FA)~\cite{Fredrickson-Andersen} and the
East~\cite{Jackle-Eisinger} spin facilitated models (see
\cite{Ritort-Sollich} for a comprehensive review). They consist of a
chain of two-state spins, $n_{i}=0,1$ $(i=1\cdots N)$, with
non-interacting Hamiltonian $H=\sum_{i} n_{i}$, and single spin flip
dynamics subject to local kinetic constraints. In the FA model, a spin
can flip if either of its nearest neighbours is in the up state, while
in the East model a spin can flip only if its nearest neighbour to the
right is up.  The equilibrium behaviour of both models is that of an
ideal gas of binary excitations, with an equilibrium concentration of
up spins given by $c=1/\left( 1+e^{1/T}\right)$ at temperature $T$. At
low temperatures the dynamics is glassy due to the competition between
decreasing the energy and the need for facilitating spins. The FA
model is Arrhenius, or strong, in the jargon of supercooled
liquids~\cite{Angell}, with relaxation time $\tr = e^{3/T}$. The East
model is super-Arrhenius, or fragile, $\tr = e^{1/(T^{2}\ln 2)}$.

\begin{figure}
\begin{center}
\epsfig{file=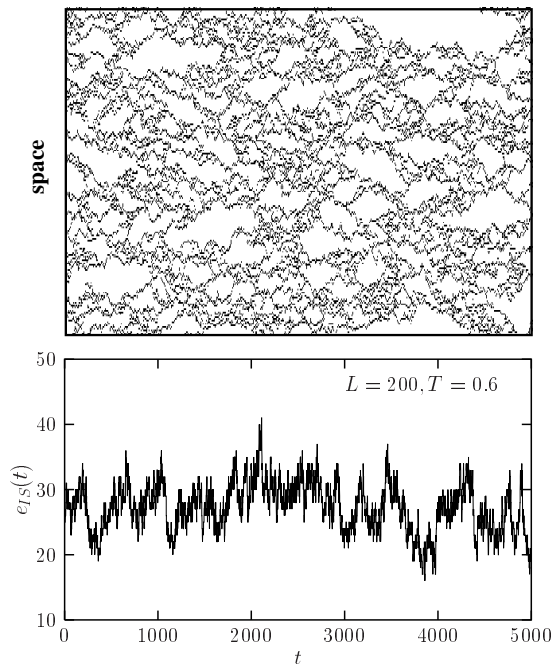, width=3.0in}
\caption{\label{ISlarge}
Top: an IS trajectory in the FA model at $T=0.6$ and system
size $L = 200 = 32 \, \ell(T)$; vertical axis is space, horizontal one
is time, up spins are black. 
Bottom: the corresponding time series of the IS energy.}
\end{center}
\end{figure}

In order to study the dynamics of IS in the FA and East models we
consider the single spin Monte Carlo dynamics in equilibrium at
temperature $T$, using a combination of Metropolis and continuous time
algorithms~\cite{Newman99}.  At each Monte Carlo step, we quench the
system by running a zero temperature Metropolis dynamics.  We thus
obtain the finite $T$ and its corresponding IS trajectory, as
initially suggested in the context of liquids. In Fig.~\ref{ISlarge}
(top), we show such an IS trajectory in the FA model at $T=0.6$. The
vertical axis corresponds to space and the horizontal one to time. Up
(down) spins in the IS configuration are denoted black (white).  The
IS structure trajectory is very close to that of the corresponding
finite $T$ one. This is due to the fact that, as a consequence of the
kinetic constraints, an up spin can be relaxed under the quench only
if a neighbour is in the up state, so that the quenching procedure
does not break the continuity of the up spin ``world-lines''.  This IS
trajectory then displays the structure of a dense mixture of ``bubbles''
separated by excitation lines, as discussed
in~\cite{Garrahan-Chandler}.  In Fig.~\ref{ISlarge} (bottom), we show
the evolution of $e_{IS}$, the (extensive) energy of the IS, as a
function of time $t$.  The time series is featureless, corresponding
to a sequence of many uncorrelated events.  This is analogous to what
is found in simulations of liquids for high temperatures or large
system sizes~\cite{SW3,Buchner-Heuer}.

\begin{figure}
\begin{center}
\epsfig{file=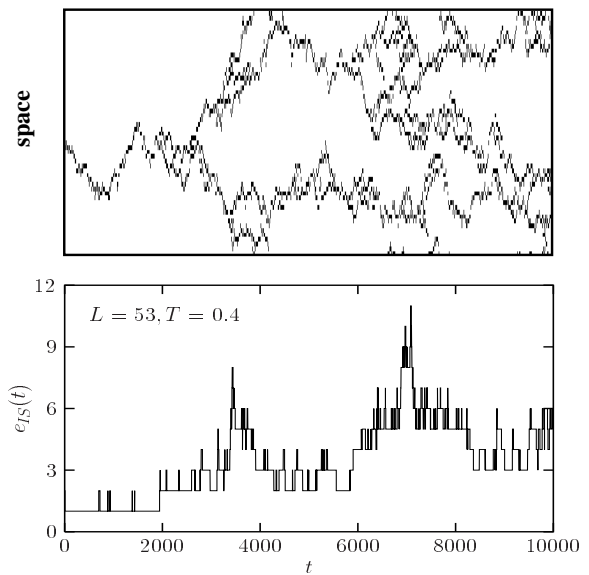, width=3.0in}
\caption{\label{ISsmall}
Same as Fig.\ 1 but for $T=0.4$ and $L=53 = 4 \, \ell(T)$.}
\end{center}
\end{figure}

The existence of a dynamical coherence length, $\ell(T)$, is however
clear from the IS trajectory in Fig.~\ref{ISlarge}.  Individual
transitions in IS correspond to the birth or closure of the trajectory
space bubbles, i.e., the branching or coalescence of excitation lines.
The dynamical correlation length $\ell(T)$ is thus given by the
typical size of the bubbles, which is fixed by the equilibrium
concentration of excitations, $\ell(T) = c^{-1}(T)$. In
Fig.~\ref{ISlarge}, the system size $L$ is much larger than $\ell(T)$,
$L = 32 \, \ell(T)$.  In Fig.~\ref{ISsmall} we show instead an IS
trajectory and its corresponding $e_{IS}$ time series in a case where
the system size is much closer to the coherence length, $L = 4
\, \ell(T)$.  The time series for $e_{IS}$ now displays the coherent
changes in IS energies also observed in numerical simulations of
``sufficiently small'' supercooled
liquids~\cite{Buchner-Heuer,Denny-et-al,SW3}.  Obviously,
``sufficiently small'' only becomes a well-defined statement when
$\ell(T)$ is first defined.  We emphasize also that these ``hidden
structures'' follow here from the presence of dynamical
heterogeneities, with no obvious relation to specific topographic
features of the energy landscape, which in this case is trivial.

All the features of the $e_{IS}$ time series in Fig.~\ref{ISsmall}
(bottom) can be traced back to the slow bubbles, i.e., dynamic
heterogeneities, in the trajectory of Fig.~\ref{ISsmall} (top). The
fast spikes in Fig.~\ref{ISsmall} (bottom) correspond to smaller
bubbles that ``wet'' the larger ones, which in turn determine the more
persistent plateaus in $e_{IS}$. In principle one could renormalize
out the smaller bubbles and obtain a time series for ``metabasins'' in
analogy with~\cite{Buchner-Heuer,Denny-et-al,Doliwa-Heuer}. 
In our case this is
unnecessary since we know the space time distribution of
bubbles~\cite{Garrahan-Chandler} and we can therefore account for the
effect in the IS dynamics of the whole range of dynamical
heterogeneities.  This is especially important in the case of the East
model, where the hierarchical nature of the dynamics leads to a
fractal wetting structure which in turn is responsible for the
temperature dependent stretching of correlation functions
characteristic of fragile
systems~\cite{Garrahan-Chandler,Buhot-Garrahan}.

\section{Lifetime of heterogeneities versus ``hopping times''}

Each event in the time series of $e_{IS}$ of Figs.~\ref{ISlarge} and
\ref{ISsmall} which increases/decreases the energy of the IS
corresponds to the birth/closure of a bubble in trajectory space. The
distribution of time intervals between events is then given by the
distribution of time extensions of bubbles, or lifetimes of dynamical
heterogeneities, $p(t)$. One has~\cite{Garrahan-Chandler}:
\begin{equation}
p(t) = \int_0^\infty \rho(\ell) \rho(t | \ell) \upd \ell
= \frac{t^{\beta-1}}{\tr^{\beta}{\cal{N}} (\beta)} 
e^{-(t/\tr)^\beta},
\label{pt}
\end{equation}
where $\rho(\ell)=c \, e^{-c \ell}$ is the distribution of length
scales of heterogeneities, $\rho(t|\ell)$ the conditional distribution
of corresponding time scales, and ${\cal N}$ a normalization
factor. The stretching exponent $\beta$ is temperature independent,
$\beta = 1/2$, for the FA model, and decreases with $T$ for the East
model, with $\beta \propto T$ at low temperatures. The correlation
functions are determined through the persistence function $P(t) =
\int_{t}^{\infty} p(t') \upd t' \simeq \exp [-(t/\tr)^{\beta}]$.
These models are an explicit realization of stretched relaxation
induced by the distribution of time and length scales of heterogeneous
regions~\cite{review}.

\begin{figure}
\begin{center}
\epsfig{file=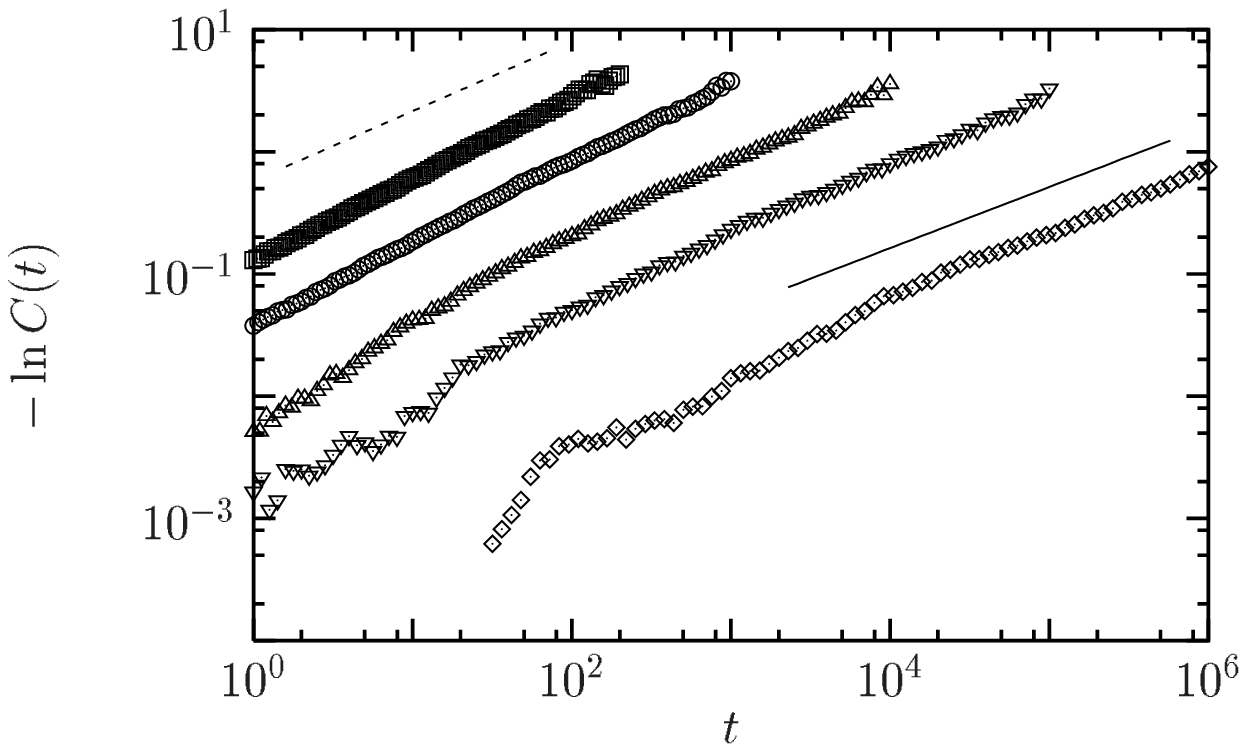, width=3.2in}
\epsfig{file=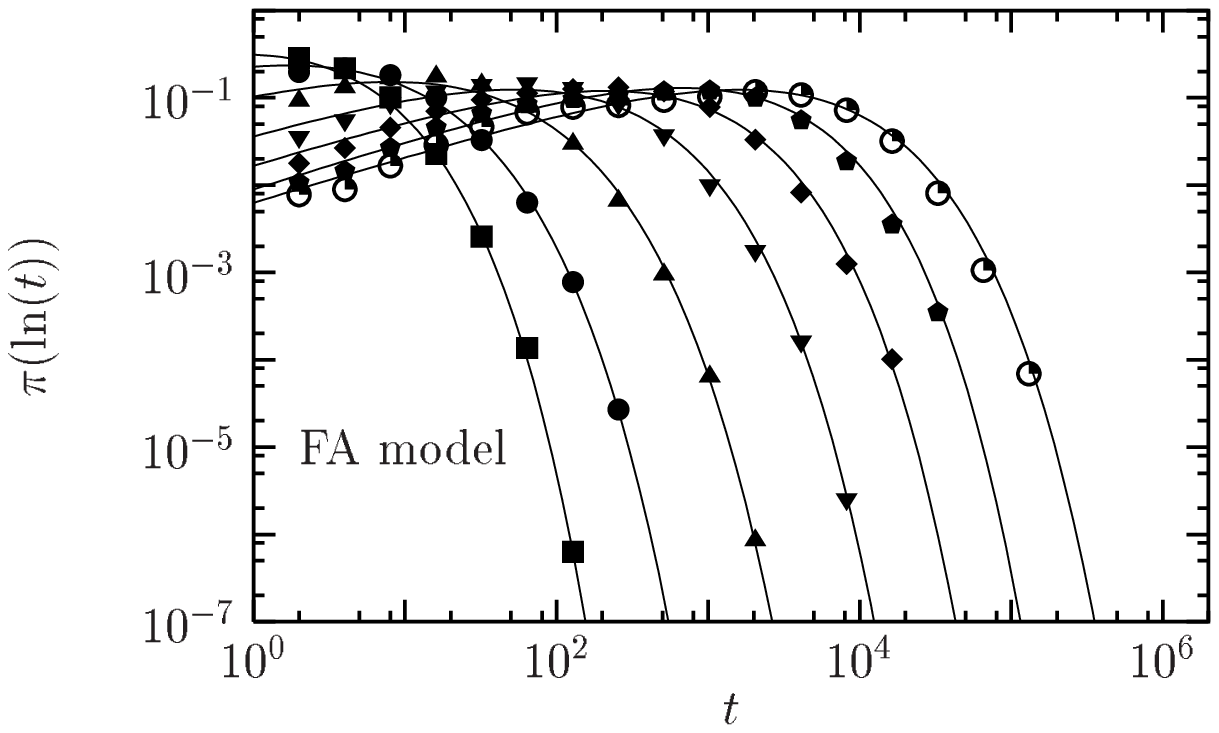, width=3.2in}
\caption{\label{fa} Top: IS energy correlation function $C(t)$ for the
FA model at various $T=1.0$, 0.6, 0.4, 0.3 and 0.2 (from left to
right) and system sizes $L = 4 \, \ell(T)$, in a log--double log
scale. The lines indicate stretching exponents $\beta=0.57$ and
$\beta=0.5$ (from left to right).  Bottom: distribution of hopping
times; symbols correspond to simulation data at temperatures $T=1.0$,
0.6, 0.4, 0.3, 0.25, 0.22, 0.2 (from left to right), and lines to fits
using Eq.~(\ref{pt}) with fixed $\beta=1/2$.}
\end{center}
\end{figure}

\begin{figure}
\begin{center}
\epsfig{file=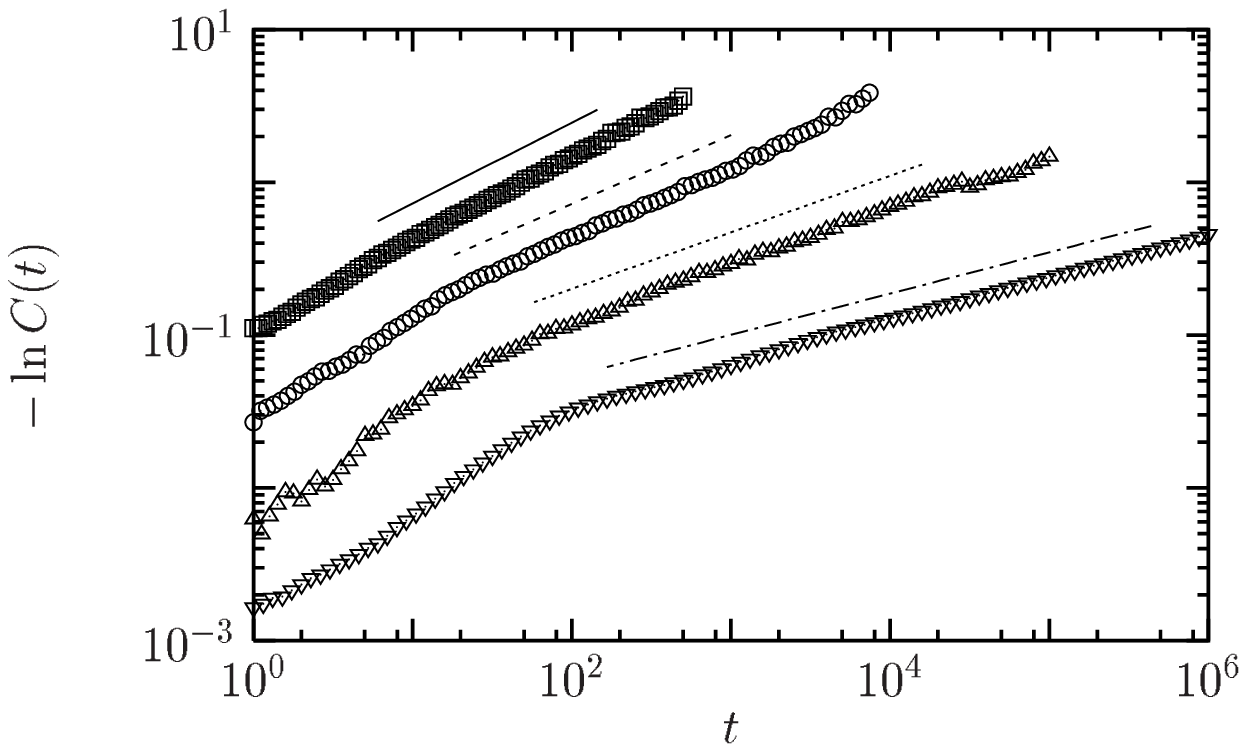, width=3.2in}
\epsfig{file=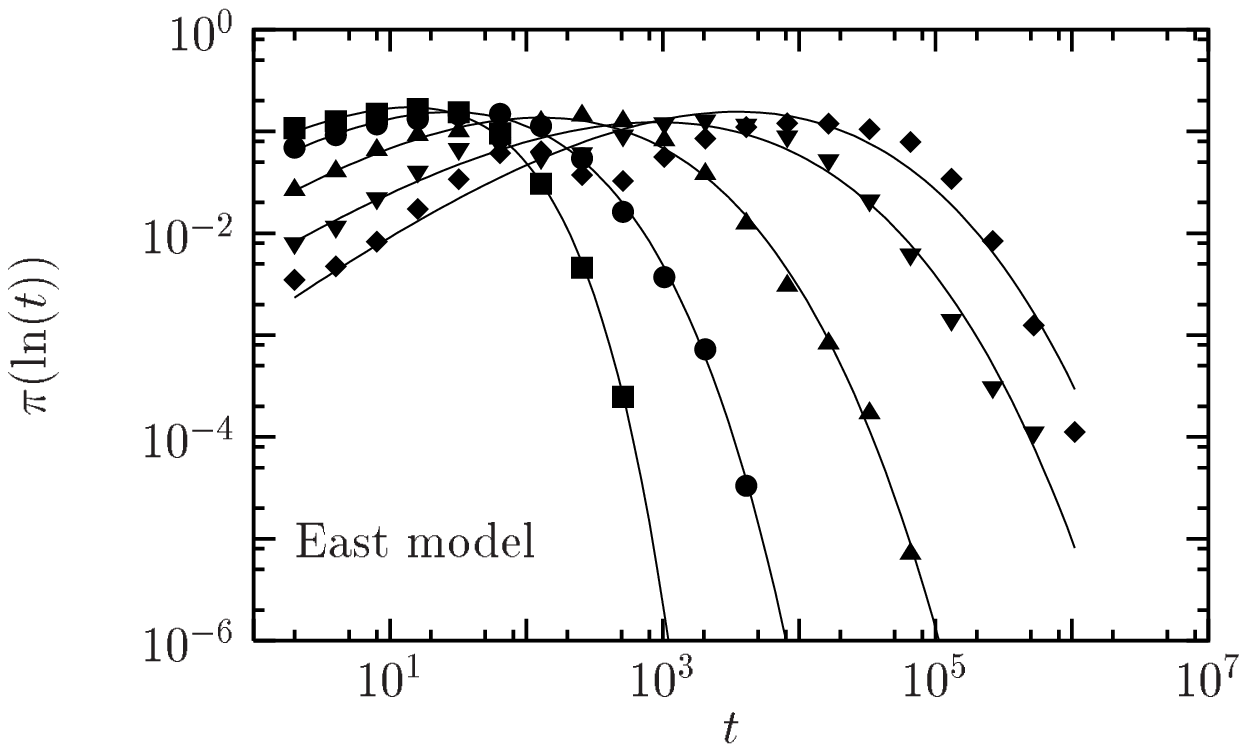, width=3.2in}
\caption{\label{east} Same as Fig.\ 3 but for East model.  Top:
Temperatures are $T=1.0$, 0.6, 0.4, 0.3 (from left to right), and the
stretching exponents are $\beta=0.55$, 0.45, 0.37 and 0.27 (from left
to right). Bottom: Temperatures are $T=1.0$, 0.6, 0.4, 0.3, 0.25 (from
left to right), the fits are from Eq.~(\ref{pit}) with $\beta$ fixed
by the correlators.}
\end{center}
\end{figure}

To numerically access the distribution (\ref{pt}), we follow the
empirical procedure of Refs.~\cite{Buchner-Heuer2,Buchner-Heuer}. As
for these simulations of supercooled liquids, we strike a compromise
between small enough system sizes, comparable to the dynamic
correlation length in order to distinguish between individual events,
with large enough ones to avoid finite size effects, corresponding
here to an upper cutoff of the integral (\ref{pt}).  Again, the
procedure becomes well-defined when $\rho(\ell)$ has first been
defined, Eq.~(\ref{pt}).  In practical terms, we seek the smallest
system size for which autocorrelation functions converge to those of
the bulk. In the case of the FA and East models we have checked that
this is obtained for systems of length $L \simeq 4 \, \ell(T)$ and $8
\, \ell(T)$, respectively. The difference stems from the hierarchical
structure of time scales of the East model which enhances the
relevance of larger length scales with respect to the FA case.

Figs.~\ref{fa} and \ref{east} show that Eq.~(\ref{pt}) indeed
describes the distribution of times between events---or ``hopping
times'' in the language of~\cite{Denny-et-al,Doliwa-Heuer}---in the IS
time series. Fig.~\ref{fa} (top) presents for the FA model, at various
low temperatures for system sizes $L=4 \, \ell(T)$, the IS energy
correlation function, $C(t) = [ \langle e_{IS}(t) e_{IS}(0) \rangle -
\langle e_{IS}^2 \rangle ] / L$.  An almost temperature independent
stretching, close to $\beta=1/2$, is obtained, as expected at low
temperatures, Eq.~(\ref{pt}).  In the lower panel of Fig.~\ref{fa} we
show the distribution of the logarithm of hopping times, $\pi(\ln t)$,
for a similar range of temperature and similar sizes. From
Eq.~(\ref{pt}), we have
\begin{equation}
\pi(\ln t) = \frac{1}{{\cal N}(\beta)}
\left( \frac{t}{\tr} \right)^{\beta} \, \exp \left[-\left(
\frac{t}{\tr} \right)^{\beta} \right] .
\label{pit}
\end{equation}
Symbols in the bottom panel of Fig.~\ref{fa} indicate simulation data,
while lines correspond to fits using Eq.\ (\ref{pit}) with
$\beta=1/2$, i.e., the single fitting parameter is the relaxation time
$\tr(T)$.  The fits are excellent over several orders of
magnitude. Fig.~\ref{east} presents a similar analysis for the East
model, for systems sizes $L = 8 \, \ell(T)$. The top panel shows
$C(t)$. The temperature dependence of the stretching is obvious, and
follows well the theoretical expectations.  The bottom panel gives the
data for hopping time distributions, and fits using Eq.~(\ref{pit}),
where now $\beta(T)$ is fixed from the corresponding values for
$C(t)$.  Again, the fits are good over several orders of magnitude.

The agreement between theory and simulations does not come as a
surprise. This step was necessary, however, to establish the actual
link between the dynamically heterogeneous dynamics of the system and
the distribution of ``hopping'' or ``trapping'' times as measured in
Refs. \cite{Denny-et-al,Doliwa-Heuer,dh2}. It also shows that these
names (hopping, trapping) somehow obscure the physical interpretation.

\begin{figure}
\begin{center}
\epsfig{file=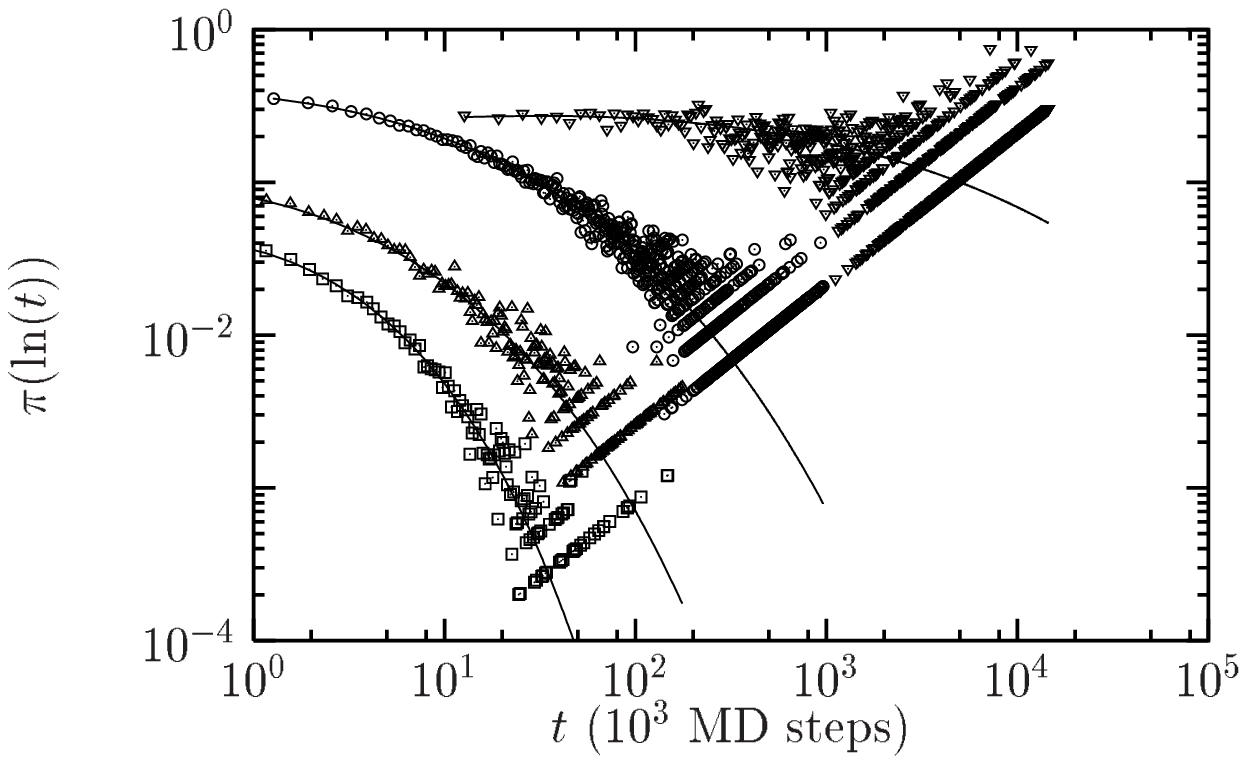, width=3.2in} \epsfig{file=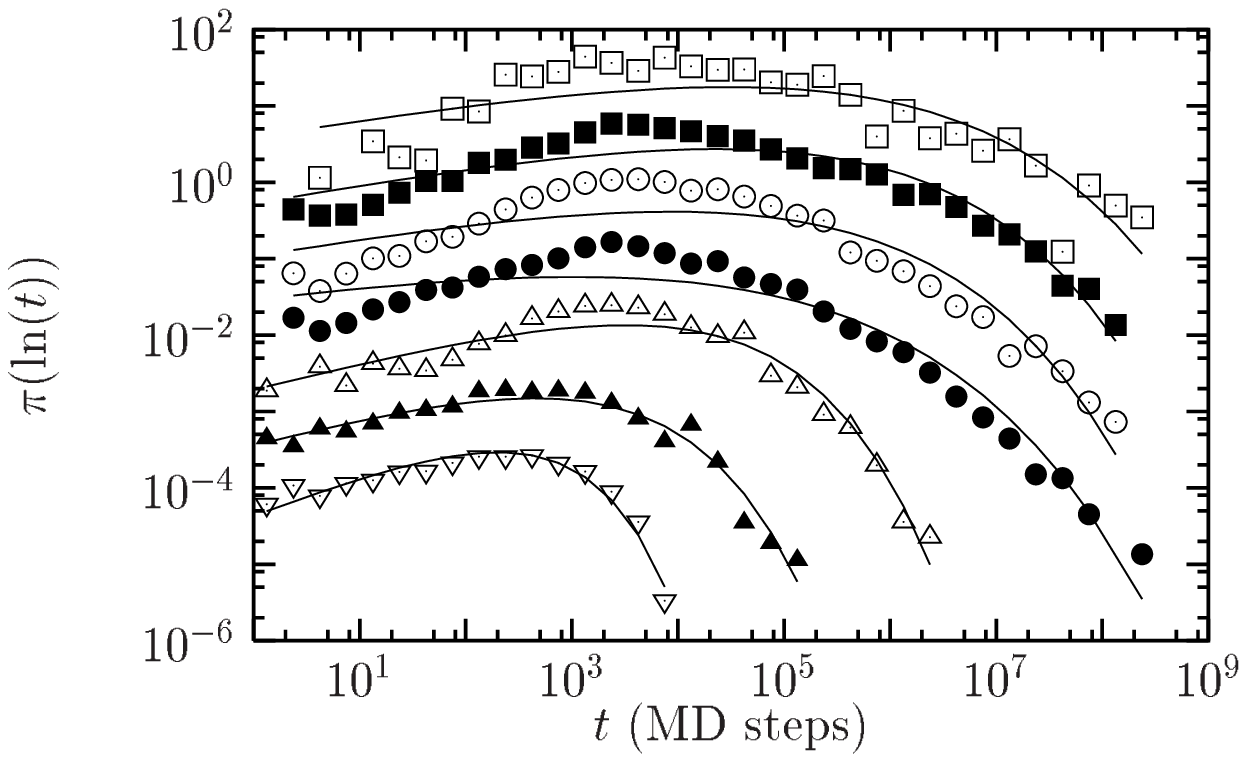,
width=3.2in}
\caption{\label{simu} Fits of LJ data from \cite{Denny-et-al} (top)
and \cite{Doliwa-Heuer} (bottom) using Eq.\ (\ref{pit}).  The points
are the published data, while the full lines are the fits, using two
fitting parameters, $\beta$ and $\tr$ at each $T$.  Top:
$(T,\beta,\tr)$ are (0.49, 0.225, 30641) (0.575, 0.275, 272) (0.669,
0.316, 257) (0.764, 0.428, 135) from top to bottom.  Bottom: (0.4,
0.234, 29500), (0.435, 0.254, 24200), (0.466, 0.246, 8820), (0.5,
0.210, 1268), (0.6, 0.358, 3138), (0.8, 0.380, 452), (1.0, 0.589, 203)
from top to bottom.}
\end{center}
\end{figure}

\section{Comparison to simulations and experiments}

The spin facilitated models studied above are a simple instance where
the phenomenon of dynamical ``trapping'' occurs due to the presence of
spatial dynamical heterogeneities. Moreover, they are simple enough
that the statistical properties of the ``traps'' can be worked out
analytically. The basic initial observation of Stillinger and Weber
was precisely the presence of these ``hidden structures in
liquid''~\cite{SW1,SW3}. More recently, quantitative studies of the
statistical properties of the IS dynamics in liquids have
appeared~\cite{Denny-et-al,Doliwa-Heuer,dh2}. The purely topographic
interpretation of the data made in
Refs.~\cite{Denny-et-al,Doliwa-Heuer,dh2} was the main motivation for
the current paper, since we think that the ``basins'', ``metabasins''
or ``traps'' described in these works are a manifestation of the
presence of dynamical heterogeneities, but not necessarily of static
``traps'', as discussed in the previous sections.

Having theoretical predictions from the spin facilitated models, it is
tempting to reanalyze published data to check whether predictions are
able to describe also more realistic models and experiments. Our
ambition here is not to claim that one-dimensional spin models are
enough to describe the physics of three-dimensional molecular
supercooled liquids, but rather to show that the mechanism described
above is a simple and robust one, with an applicability well beyond
these simple models \cite{arrows}.

\begin{figure}
\begin{center}
\epsfig{file=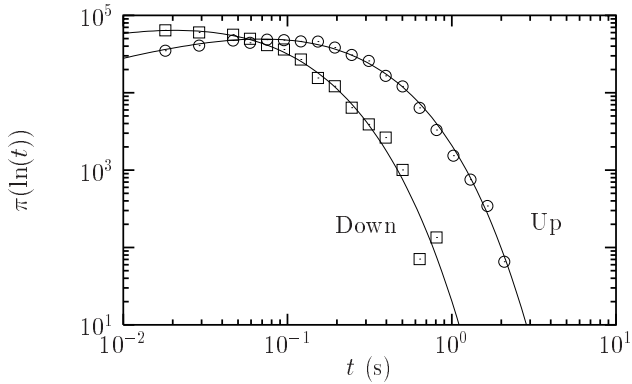, width=3.5in}
\caption{\label{isr}
Distributions of molecular switching times in polyvinyl
acetate at $299$ K, from Ref.~\cite{VidalRussell-Israeloff}, and fits
using Eq.~(\ref{pit}) with $\beta=0.67$, $\tr=0.070$ s (Up), and
$\beta=0.63$, $\tr=0.021$ s (Down).}
\end{center}
\end{figure}

In Fig.~\ref{simu}, we present the data for hopping time distributions
from the simulations of~\cite{Denny-et-al} (top)
and~\cite{Doliwa-Heuer} (bottom). The lines through the symbols are
fits using the distribution for dynamical heterogeneities log times of
the FA/East models, Eq.\ (\ref{pit}). Apart from an irrelevant
normalization constant, there are only two fitting parameters, the
stretching exponent $\beta$ and the relaxation time $\tr$. The fits to
the data of Denny et al. (top) are as good or better than those with a
log normal distribution, as predicted by an activated
scaling~\cite{Denny-et-al}.  The agreement is also very good for
Doliwa and Heuer's data (bottom), which we have chosen to present as
$\pi(\ln t)$ rather than as $p(t)$ as they were originally published
in~\cite{Doliwa-Heuer}.  The deviations in the fits in this case are a
consequence of the data for the very short times investigated
in~\cite{Doliwa-Heuer}, where we expect the mechanism of dynamical
facilitation to be less relevant, so that Eq.~(\ref{pit}) should
describe the tails of the distributions more accurately than the small
times~\cite{fit}.

Finally, we have compared Eq.~(\ref{pit}) to experimental
observations. Although changing the system size is not conveniently
realized in experiments, it is still possible to investigate the
dynamics in ``mesoscopic'' regions, i.e., regions of a size comparable
to the dynamic coherence length $\ell(T)$.  In
Ref.~\cite{VidalRussell-Israeloff}, time series for molecular
polarization were obtained by Vidal Russell and Israeloff in local
probe experiments on polyvinyl acetate films. Typical length scales
probed through the use of non-contact atomic force microscopy
techniques are $\simeq 20~{\rm nm}$, i.e., not very large compared to
the estimated size of heterogeneities~\cite{review}.  Fortunately,
since the polarization signal oscillates between two or four discrete
levels, fast thermal fluctuations only produce oscillations about
these discrete values, so that the corresponding distributions for
switching times can directly be compared to Eq.~(\ref{pit}), therefore
bypassing the IS construction.

We reproduce in Fig.~\ref{isr} the distributions presented in the
Fig.~5 of Ref.~\cite{VidalRussell-Israeloff}, where we have adopted a
log scale for the time axis, and represented $\pi(\ln t)$ instead of
$p(t)$, so that the comparison with the theoretical prediction
(\ref{pit}) can be made on the entire range of time scales and also to
facilitate the comparison with the results presented above.  Again, we
obtain an excellent fit of the experimental data to Eq.~(\ref{pit})
for the complete experimental time window.  The stretching exponents
we find, $\beta=0.67$ and $0.63$ for the up and down polarizations,
respectively, are slightly larger than the ones estimated in
\cite{VidalRussell-Israeloff}. This is because the latter stemmed from
a fit to a stretched exponential form of the tail of the
distribution. Instead, it is clear from Fig.~\ref{isr} that the power
law prefactor is necessary to account for the whole spectrum of time
scales, and this shifts the stretching exponent to a slightly larger
value.

\section{Conclusions}

We have shown that a consistent interpretation of the IS dynamics can
be obtained solely invoking spatial and dynamical aspects related to
the statistical properties of the dynamic heterogeneity of supercooled
liquids. This physical picture was made more precise in the study of
two simple dynamically facilitated spin models, which also allowed to
obtain quantitative predictions which compare well to numerical and
experimental data. Our results are additional evidence that dynamical
heterogeneities play a crucial role in the physics of liquids
approaching their glass transition~\cite{review}.

In the topographic language, the so-called ``landscape-influenced''
regime evidenced by a decrease of $\langle e_{IS} \rangle$ with
decreasing temperature is the relevant one in numerical
simulations~\cite{topo,sastry}. In our non-topographic perspective,
this decrease of the IS energy corresponds to a decreasing
concentration of facilitating defects~\cite{arrows}, i.e., to the
growth of the dynamical coherence length which is in turn directly
responsible for the dynamical slowdown. 
It is therefore striking that the physics at relatively high temperatures,
i.e., above the estimated mode--coupling temperature $T_c$, can be
described by the same physical mechanism of dynamic facilitation 
initially supposed
to be relevant close to the experimental glass transition $T_g$.
Hence, our results imply that the dynamics above $T_c$ is 
heterogeneous, in agreement with simulations~\cite{simuhetero}. 

Moreover, as suggested in~\cite{Denny-et-al,Doliwa-Heuer}, it 
logically follows that the
physics in this regime is qualitatively 
not very different from the one
at temperatures closer to $T_g$. This is at odds with the
common belief that a change of mechanism takes place close to 
$T_c$~\cite{Debe}. 
This raises two interesting questions that we 
formulate as a conclusion to the paper.
First, in the language of heterogeneities, 
what is the meaning, if any, of the 
crossovers reported close to $T_c$? 
Second, since these crossovers somehow justify 
the use of the mode-coupling theory above $T_c$, why 
does the theory apparently work in a regime where the dynamics
is heterogenous, and why does it eventually break down at lower 
temperatures?

\begin{acknowledgments}
We are grateful to J.-P. Bouchaud, D. Chandler, D.R. Reichman and
G. Tarjus for discussions.  We thank the authors of
Refs.~\cite{Denny-et-al,Doliwa-Heuer,VidalRussell-Israeloff} for
kindly providing their published data.  We acknowledge financial
support from a European Marie Curie Fellowship No HPMF-CT-2002-01927,
CNRS (France), Worcester College Oxford, EPSRC Grant No.\
GR/R83712/01, and the Glasstone Fund.  Some of the numerical results
were obtained on Oswell at the Oxford Supercomputing Center, Oxford
University.
\end{acknowledgments}


\begin{thebibliography}{99}
\bibitem{fit} We have checked that by making the exponent in the
prefactor of (\ref{pit}) independent of $\beta$, thus allowing for an
extra fitting parameter, we get excellent fits to the data.  We have
chosen not to present these fits, since our aim is less to propose a
fit of numerical results than to discuss their physical
interpretation.
\end{thebibliography}
\end{document}